\documentclass[10pt,aps,prl,twocolumn,superscriptaddress,showpacs]{revtex4}
\usepackage[final]{pdfpages}
\usepackage{graphics}
\usepackage{epsfig}
\usepackage{dcolumn}
\usepackage{bm}
\usepackage{amsmath}
\usepackage{amsfonts}
\usepackage{latexsym}
\usepackage{amssymb}
\usepackage{color}
\usepackage{hyperref}
\usepackage[normalem]{ulem}

 \newcommand{\tb}[1]{}

\newcommand{\g}[1]{{\bf #1}}

\newcommand{\yb}{YbB$_{12}$}
\newcommand{\sm}{SmB$_6$}
\newcommand{\cebipt}{Ce$_3$Bi$_4$Pt$_3$}
\newcommand{\crs}{CeRhSb}
\newcommand{\cns}{CeNiSn}

\newcommand{\eqn}[1]{(\ref{#1})}
\newcommand{\be}{\begin{equation}}
\newcommand{\ee}{\end{equation}}
\newcommand{\bea}{\begin{eqnarray}}
\newcommand{\eea}{\end{eqnarray}}
\newcommand{\ba}{\begin{eqnarray*}}
\newcommand{\ea}{\end{eqnarray*}}
\newcommand{\dagga}{{\phantom{\dagger}}}
\newcommand{\up}{\uparrow}
\newcommand{\down}{\downarrow}
\newcommand{\dis}{\displaystyle}
\newcommand{\fract}[2]{\frac{\dis #1}{\dis #2}}

\begin{document}

\title{Many-body breakdown of indirect gap in topological Kondo insulators}

\author{Marcin M. Wysoki\'nski}
\email{mwysoki@sissa.it}

\affiliation{International School for Advanced Studies (SISSA)$,$ via Bonomea 265$,$ IT-34136$,$ Trieste$,$ Italy}
\affiliation{Marian Smoluchowski Institute of Physics$,$ Jagiellonian
University$,$ 
ulica prof. S. \L ojasiewicza 11$,$ PL-30-348 Krak\'ow$,$ Poland}

\author{Michele Fabrizio}
\email{fabrizio@sissa.it}
\affiliation{International School for Advanced Studies (SISSA)$,$ via Bonomea 265$,$ IT-34136$,$ Trieste$,$ Italy}

\date{\today}

\begin{abstract}
We show that the inclusion of nonlocal correlation effects 
in a variational wave function for the ground state of a 
topological Anderson lattice Hamiltonian is capable 
of describing both topologically trivial insulating phases and 
nontrivial ones characterized by an indirect gap, as well as its  
closure at the transition into a metallic phase. 
The method, though applied to an oversimplified model, 
thus captures the metallic and insulating 
states that are indeed observed in a variety of 
Kondo semiconductors, while accounting for topologically nontrivial band structures.   
 \end{abstract}

 

\maketitle

{\it Introduction.}\
Phenomena at the crossroads of  
topological insulators and strongly 
correlated systems have recently gained a lot of attention,  
mainly stimulated by theoretical 
proposals that the inclusion of strong correlations in 
specific models with a nontrivial 
topological content may give rise to 
novel and interesting phenomena 
\cite{Assaad_rev,Raghu2008,Kivelson2009,Sarma2012,Amaricci2015}. 
However, electronic correlations in most 
of the discovered topological insulators \cite{RMP_TI2,RMP_TI1} seem to be relatively weak, and hence of minor importance.
For that reason,  the concept of topological Kondo insulators (TKIs) originally put forward in Refs.~\cite{Coleman2010} and \cite{Coleman2012} is particularly appealing 
due to its possible realization in already known 
Kondo insulating compounds \cite{Riseborough2000}. 
Notably, \sm\ is convincingly confirmed to be a TKI 
\cite{Durakiewicz2013,Jiang2013,Fisk2013,Ruan2014,Fisk2014,Fisk_nat,
Syers2015,Thompson2015,Thompson2015_1}
with strong evidence of the essential role played by many-body correlations
\cite{Dai2013,Fuhrman2015}.

Theoretically, the main physical properties of 
TKI are frequently described in the framework of 
topological Anderson lattice models in which 
strong spin-orbit coupling is encoded into 
a spin-dependent hybridization with odd parity 
in momentum space \cite{Coleman2010,Coleman2012,Kim2012}. 
Many-body correlations in these models have been mostly treated
by the slave-boson approach   
or by the dynamical mean-field theory (DMFT),   
and predicted to induce 
quantum phase transitions between topologically distinct 
bulk insulating phases \cite{Kim2012,Assaad2013, Sigrist2014}.

The aim of this  work  is to demonstrate that accounting 
for nonlocal spatial correlations, 
beyond those already well captured by the mentioned techniques,  
plays an important role in modeling TKIs. Specifically, those 
nonlocal correlations supply the $f$ electron self-energy with a momentum dependence, which is otherwise purely local within DMFT and at the saddle point of the slave-boson theory. Remarkably, such a momentum dependence allows one to describe, above a critical interaction 
strength, the emergence of a topological Kondo insulator with an indirect gap, and its subsequent closure at 
the transition into a metallic state, 
an intriguing result {\it opposite} to 
the conventional Mott phenomenon where increasing 
interaction instead favours the onset of an insulating state.  
We disclose this scenario in two dimensions by means of the 
diagrammatic expansion of the Gutzwiller wave function (GWF)
technique \cite{buenemann2012,Wysokinski2015}. 

 The emergence of an indirect gap \cite{Menth1969} or its closure \cite{Takabatake1990} from 
a model of nondispersive $f$ orbitals, in accordance with their large mutual distance in actual materials \cite{Butch2016,Hiess1997}, 
is relevant to candidate TKIs, e.g., \sm, 
but also to other $4f$  compounds not yet excluded to host topologically nontrivial states, which show bulk insulating behavior (\cebipt\ \cite{Hundley1990}, \yb\ \cite{Okamura1998}, CeRhAs\- \cite{Yoshii1996}) as well as metallic one
(\cns\ \cite{Takabatake1990},\crs\ \cite{Malik1991}, CeIrSb \cite{Takabatake2007}). Moreover, assuming that increasing pressure roughly corresponds 
to reducing the interaction strength, we can qualitatively account 
for the nonuniversal behavior of the gap as observed in several 
Kondo semiconductors \cite{Hiraoka1994,Yoshii1996,Cooley1997,Fisk1996,Bauer2003,Fisk2016}. 

{\it Model and method.}\
 Our starting point is the topological Anderson lattice model on a square lattice,
\begin{eqnarray}
 \mathcal{H}\!\!\!&=&\!\!\!\!\sum_{\g i,\g j}\, 
 t_{\g i\g j}\; \hat c_{\g i}^\dagger\hat c^\dagga_{\g j}
-\!\sum_{\g i}\bigg(\mu\,\hat c_{\g i}^\dagger\hat c^\dagga_{\g i}+ 
\big(\mu-\varepsilon_f\big)\,\hat f_{\g i}^\dagger\hat f^\dagga_{\g i}+
U \,  n_{\g i\uparrow}^f  n_{\g i\downarrow}^f\Big)\nonumber \\
&& +\!\!\!\!\
\sum_{\langle\g i,\g j\rangle_\alpha,\alpha=x,y}\!\!\!\bigg(iV\,\Big(\hat c_{\g i}^\dagger\,\sigma_\alpha\,
\hat f_{\g j}^\dagga+\hat f_{\g i}^\dagger\,\sigma_\alpha\,
\hat c_{\g j}^\dagga\Big)+\rm{H.c.}\bigg),
 \label{ALM}
\end{eqnarray}
where the two-component spinors $\hat f_{\g i}^\dagger\equiv \big(\,f_{\g i\uparrow}^\dagger\,,\, 
f_{\g i\downarrow}^\dagger\,\big)$, and accordingly for $\hat c_{\g i}^\dagger$,  
describe localised ($f$) and conduction ($c$) electrons, $\sigma_{(\alpha=x,y)}$ are the Pauli matrices, and $\langle\g i, \g j\rangle_{\alpha}$ denote 
pairs of nearest-neighbor sites in the $\alpha=x,y$ direction. 
The nontrivial topology resulting from spin-orbit coupling is encoded
in the odd-parity hybridization between $f$ and $c$ electrons \cite{Coleman2010,Coleman2012,Kim2012}.  
Throughout this work we shall assume nearest neighbor $c$--$c$ hopping ${t\!=\!-1}$, and $c$--$f$ hybridization $V\!=0.5$. 

We study the model Eq.~\eqn{ALM} with a variational GWF ${| \psi_G \rangle\equiv\mathcal{ P}_G|\psi_0 \rangle\equiv \prod_{\g i}
\mathcal{P}_{\g i}\,| \psi_0 \rangle}$ constructed from a Slater determinant, $|\psi_0\rangle$ modified 
by the application of local linear operators $\mathcal{ P}_{\g i}$ defined through~\cite{Gebhard1990} 
\begin{equation}
\begin{split}
\mathcal{P}_{\g i}^\dagger\, \mathcal{P}^\dagga_{\g i}=\mathcal{ P}_{\g i}^2&\equiv{\bf1} +x\, 
d_{\g i},
\label{varx}
\end{split}
\end{equation} 
where $x$ is a variational parameter, 
\be
d_{\g i} = \Big(n^f_{\g i\up} - \langle n^f_{\g i\up}\rangle_0\Big)
\Big(n^f_{\g i\down} - \langle n^f_{\g i\down}\rangle_0\Big),\label{def:d}
\ee
and hereafter $\langle \dots \rangle_0$ and $\langle \dots \rangle_G$ shall denote the normalised
averages over the Slater determinant, $|\psi_0\rangle$, and the GWF, $|\psi_G\rangle$, respectively. 
Eq.~(\ref{varx}) allows one to write the exact expectation value of any operator $\mathcal{O}_I$ 
that involves sites $I=\big\{\g i_1,\g i_2, \dots,\g i_M\big \}$
as a power series in the parameter $x$,
\begin{equation}
\begin{split}
 \big\langle \, &\mathcal{O}_{I}\,\big\rangle_G = \frac{1}{\langle\psi_G|\psi_G\rangle}
 \Big\langle \,{\mathcal{O}}^G_{I}\;\prod_{\g l\not \in I}\,  
 \mathcal{P}_{\g l}^2 \, \Big \rangle_0    \\
 &= \frac{1}{\langle\psi_G|\psi_G\rangle}\sum_{k=0}^\infty\, \fract{x^k}{k!}\, \sum_{\g l_1,...,\g l_k}\!\!\!\!{}^{'}
 \Big\langle \,{\mathcal{O}}^G_{I}\, d_{\g l_1}\,d_{\g l_2}\dots d_{\g l_k}\,\Big
\rangle_0\; ,
 \label{expe}
\end{split}
 \end{equation}
where  $\mathcal{O}^G_{I}\equiv \prod_{i\in I}\,\mathcal{P}_{\g i}^\dagger\;\mathcal{O}_I\; 
\prod_{j\in I}\,\mathcal{P}_{\g j}^\dagga$. 
The primed summation is restricted to sites $\not\in I$. 
The expectation values in Eq. (\ref{expe}) are 
evaluated by means of the Wick's theorem.

The popular Gutzwiller approximation corresponds to keep just the $k=0$ term in 
Eq.~(\ref{expe})~\cite{Wysokinski2015}, which is in fact the only term that survives 
in the limit of infinite dimensions. In this simple case the variational optimization 
reduces to fix a single parameter, namely, the double occupancy of $f$ electrons.
To account for nonlocal spatial correlations in finite dimensions,  
higher orders of the expansion ($k>0$) are systematically incorporated in  
Eq.~\eqn{expe} and expectation values of the products of the operators are evaluated diagrammatically \cite{buenemann2012,Wysokinski2015,
Kaczmarczyk2013,Kaczmarczyk2014,
Tomski2016,Kaczmarczyk2014P,Kaczmarczyk2015,SC_arxiv}.  By construction~\cite{buenemann2012}, the
convergence of the sum with respect to $k$ is reached relatively quickly, and 
here we found that $k=4$ is already satisfying (see the Supplemental Material \cite{supl}). 
We emphasize  that the convergence is not related to $x$ being a small expansion parameter, but rather to the fast decrease of the expectation values at subsequent orders.
For numerical tractability, we must set a real-space cut-off distance 
beyond which we neglect in Eq.~\eqn{expe} the contribution of the nonlocal components 
of the single-particle density matrix 
$C_{\g{r},\g{0}}^{\alpha,\beta}\equiv \langle \alpha_{\g{r}\sigma}^\dagger
  \beta_{\g{0}\sigma'}^\dagga \rangle_0$, where $\alpha,\beta \in\{f,c\}$,  $\sigma=\sigma'$ for $C^{ff}$ and $C^{cc}$, and $\sigma=-\sigma'$ for $C^{cf}$. Specifically we choose 
$|\g r|^2\leq5$ for $C^{ff}$ and $|\g r|^2\leq2$ for $C^{cf}$ (in units of the lattice constant).

It is known that a faithful variational description of correlated systems in finite dimensions requires going beyond the simple GWF by adding longer-range correlations via Jastrow-like operators \cite{Capello2005,Tocchio2016}, 
whose optimization can, however, be accomplished only by variational Monte Carlo techniques. Alternatively, one could be satisfied with 
just the GWF result Eq.~\eqn{expe} at higher orders in $k>0$    
on the provision that 
the uncorrelated Slater determinant is also modified accordingly 
so as to minimize the total energy. This approach, though less accurate, is numerically less demanding and provides results already in the thermodynamic limit.
Despite the simple form of the variational GWF, 
this method frequently yields results in accordance with 
the variational Monte Carlo \cite{Kaczmarczyk2013,Kaczmarczyk2014}.  
 
The expectation value of the Hamiltonian (\ref{ALM}),
 $\langle\mathcal{H}\rangle_G$ is evaluated diagrammatically according to 
Eq.~(\ref{expe}), and subsequently optimized
with respect to the variational parameter $x$ and to the single-particle Hamiltonian whose ground state 
is $|\psi_0\rangle$ (for details, see the Supplemental Material \cite{supl}), 
which turns out to have the general expression 
\begin{eqnarray}
\mathcal{H}^{\rm eff}\!\!\!\!&=& \sum_{\g i,\g j}
 t_{\g i\g j} \, \hat c_{\g i }^\dagger\, \hat c_{\g j }^\dagga
+\sum_{\g i,\g j}t_{\g i \g j}^{f}\, \hat f_{\g i }^\dagger \hat f_{\g j}^\dagga\,
\nonumber \\
&&\!\!\!\!\!\!\!+\!\!\!\sum_{\langle\g i,\g j\rangle,\alpha=x,y}\bigg[\,iV_{\g i \g j}^{\rm eff}\,\Big(\hat c_{\g i}^\dagger\,\sigma_\alpha\,
\hat f_{\g j}^\dagga +\hat f_{\g i}^\dagger\,\sigma_\alpha\, 
\hat c_{\g j}^\dagga\Big) +\rm{H.c.}\,\bigg]\label{heff}\\
&&\!\!\!\!\!\!\!\!\!\!\!\!\!\!+\!\!\!\sum_{\langle\langle\g i,\g j\rangle\rangle}\bigg\{iV_{\g i \g j}^{\rm eff}\Big[\hat c_{\g i}^\dagger\big(\sigma_x-a\sigma_y\big)
\hat f_{\g j}+\hat f_{\g i}^\dagger\big(\sigma_x-a\sigma_y\big)
\hat c_{\g j}\Big]+\rm{H.c.}\bigg\}\nonumber\\
&\equiv& \sum_{\mathbf{k}}\, \hat{\Psi}_\mathbf{k}^\dagger\, \mathcal{\hat{H}}^\text{eff}(\mathbf{k})\,
\hat{\Psi}_\mathbf{k}^\dagga,\nonumber
\end{eqnarray}
where $\hat{\Psi}_\mathbf{k} = \big(\hat{c}_\mathbf{k}\,,\,\hat{f}_\mathbf{k}\big)$ are four-component spinors in momentum space, 
$t_{\g i\g i}$ and $t^f_{\g i\g i}$ are the on-site energies of the $c$  and $f$ electrons, respectively, 
$\langle\langle\g i,\g j\rangle\rangle$ denotes the sum over next-to-nearest neighbors, and by convention, 
$a=\pm 1$ for bonds in the $(1,\pm 1)$ direction. 

In other words, the optimization of inter-site correlations in Eq.~\eqn{expe} leads to an effective single-particle Hamiltonian \eqn{heff} that includes further neighbors $c$--$f$ hybridization as well as a direct long-range $f$--$f$ hopping. On the contrary, since $\mathcal{P}_G$ does not act on the $c$  electrons, the nearest-neighbor $c$--$c$ hopping amplitude is that of the original Hamiltonian.  

The topological properties of the model may be inferred already from the 
effective Hamiltonian $\mathcal{\hat{H}}^\text{eff}(\mathbf{k}^*)$, Eq.~\eqn{heff}, which at the time reversal invariant momenta, $\g k^*$, can be casted into the form \cite{Coleman2010}, 
\begin{eqnarray*}
2\,\mathcal{\hat{H}}^\text{eff}(\mathbf{k}^*) \!=\! \Big[\epsilon_c\big(\g k^*\big)+\epsilon_f\big(\g k^*\big)\Big] \,\mathbb{I}+\Big[\epsilon_c\big(\g k^*\big)-\epsilon_f\big(\g k^*\big)\Big]\,\mathbb{P}\,, 
\end{eqnarray*}
where $\mathbb{I}$ is the identity and $\mathbb{P}$ the diagonal parity operator with elements $+1$ for $c$ and $-1$ for $f$, while the $c$ and $f$ dispersions,     
$\epsilon_c(\g k)$ and $\epsilon_f(\g k)$, are derived from Eq.~(\ref{heff}).
The $Z_2$ topological invariant $\nu$ \cite{Kane2005} can be readily obtained \cite{FuKane} from $(-1)^\nu=\prod_i\delta_i$ with 
$\delta_i=\text{sgn}\big[\epsilon_c(\g k^*_i)-\epsilon_f(\g k^*_i)\big]$, where the product runs over the four $\g k^*$-points: $\Gamma$, X, Y and M. By symmetry X and Y are equivalent.
 
\begin{center}\vspace{-0.3cm}
  \begin{figure}[t]   
\includegraphics[width=0.48\textwidth]{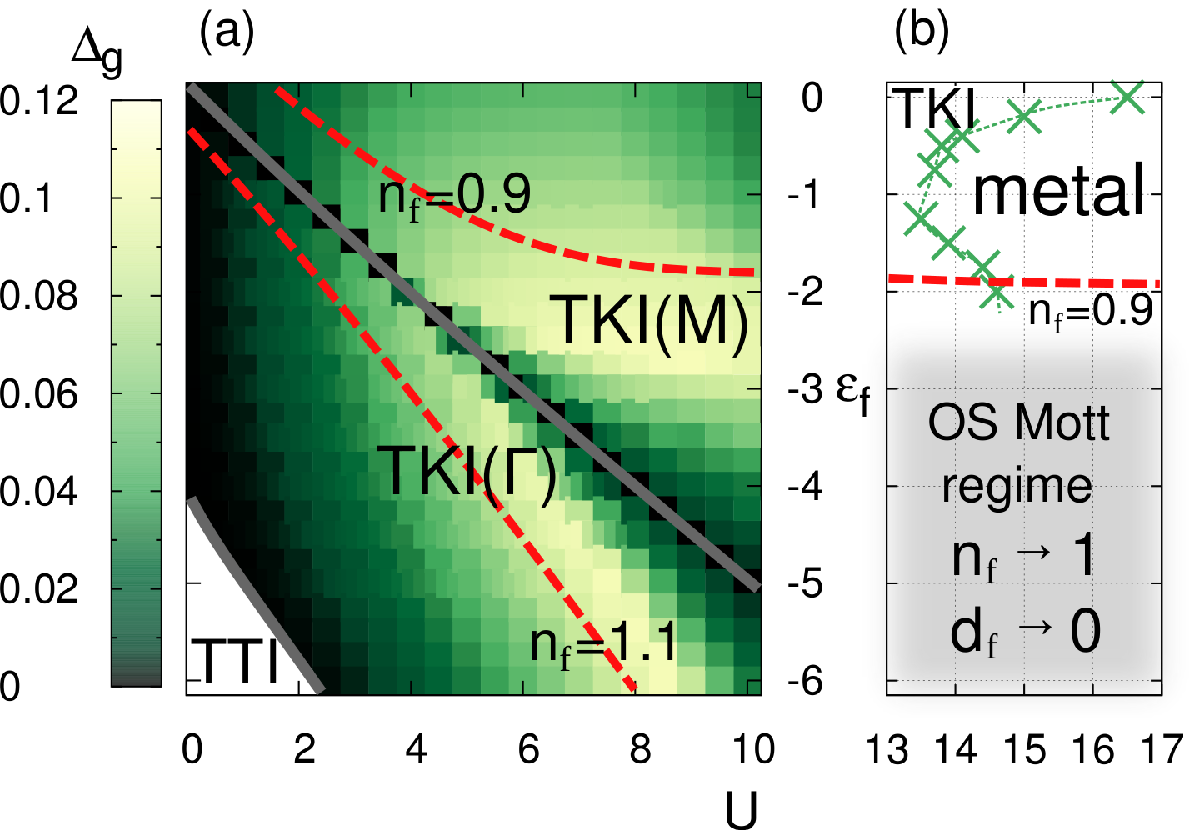}   
\includegraphics[width=0.48\textwidth]{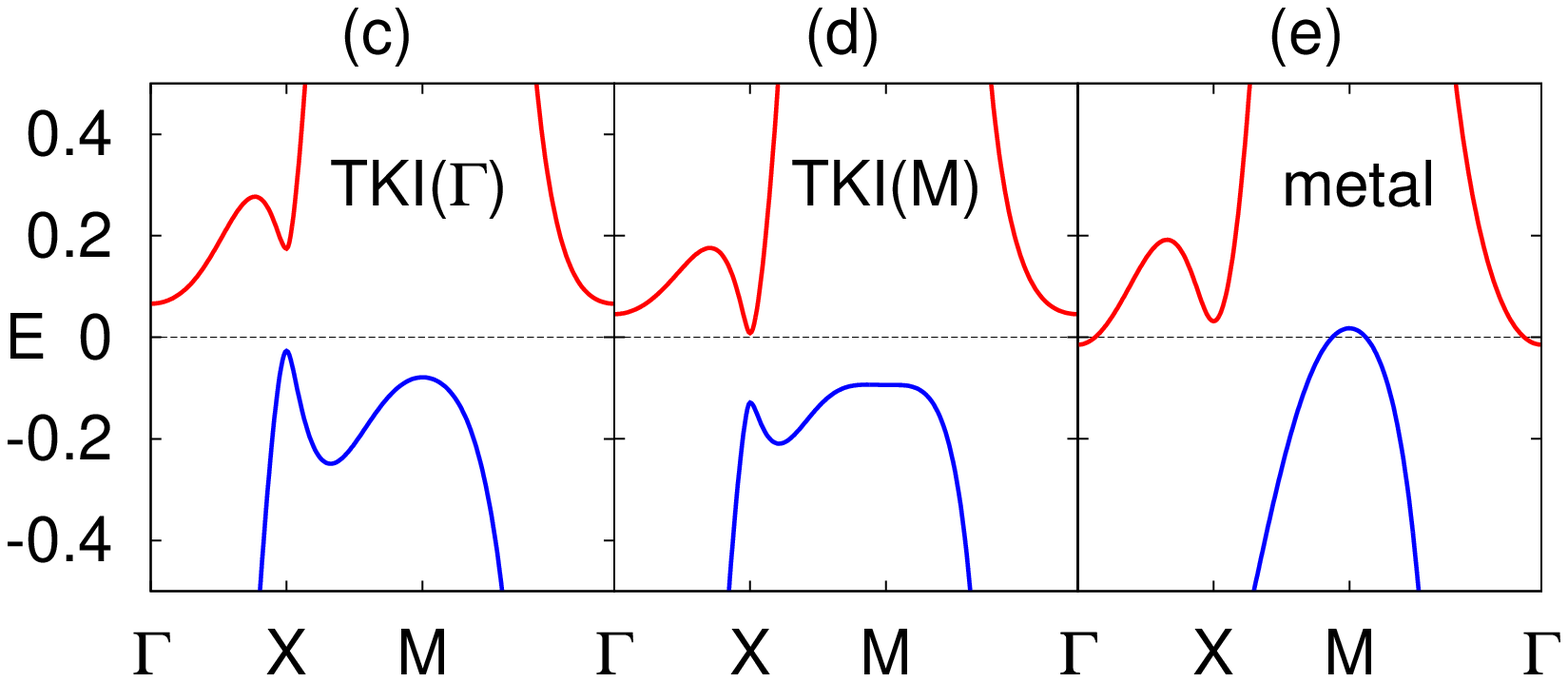}   
   \caption{(a) $U$ vs. $\varepsilon_f$ phase diagram of the model \eqn{ALM} at half-filling, with the value of the emergent indirect gap marked on a color scale. The phase diagram comprises three topologically distinct phases, a trivial topological insulator (TTI) and two nontrivial ones, TKI($\Gamma$) and TKI(M).  
Along the two red dashed lines the $f$ occupancy is constant, with $n_f=0.9$ and $n_f=1.1$. (b) The continuation of the phase diagram for larger $U$ where the TKI can undergo a transition to a metallic state. 
The region where the $f$ occupancy locks at $n_f=1$ with negligible fluctuations, $d_f\to 0$, is denoted as an orbital selective (OS) Mott regime.
(c-e) Exemplary low energy band structures for the selected phases: 
(c) TKI($\Gamma$) for $U=7$ and $\varepsilon_f=-5$; (d) TKI(M) for $U=9$ and $\varepsilon_f=-2.5$; (e) metal for $U=20$ and $\varepsilon_f=-0.5$. The valence and the conduction bands are marked with different colors. \vspace{-0.4cm}} 
  \label{fig1}
  \end{figure} 
 \end{center}\vspace{-0.7cm}

{\it Results.}\ 
We mentioned that the model, Eq.~(\ref{ALM}), with nondispersive $f$ states should hopefully describe putative TKIs such as \sm\ 
with indirect gaps between the valence and conduction bands \cite{Menth1969,Taylor2006}. However, the mean-field treatment of  
the Hamiltonian \eqn{ALM} at half filling leads instead to a semimetal because of the vanishing hybridization at $\g k^*$. 
By contrast, mean-field theory applied to models for Kondo insulators with even-parity hybridization does lead to the desired insulating state. 
To cure the inadequacy of mean-field and recover an insulator with an indirect gap, \textsl{ad hoc} dispersion of $f$ electrons 
is frequently assumed \cite{Sigrist2014,Assaad2013}, whose minimum in the Brillouin zone is shifted by ($\pi,\pi$) from that of conduction electrons. 
However, suchan  additional ingredient is not truly justified, e.g., in \sm, where the large separation between $4f$ elements  \cite{Butch2016}, hence their negligible mutual overlap, implies that the main 
source of $f$ itineracy remains the $c$--$f$ hybridization.
We mention that spin-polarized local density approximation (LSDA)+$U$ calculations \cite{LSDA+U} suggest a sizable $f$-$f$ hopping mediated by the hybridization with boron $p$ states. We suspect this is rather an artifact of the method that artificially pushes spin-majority $f$ states down to the occupied boron $p$ bands. In reality, such boron mediated hopping should be fairly negligible at the Fermi level.  

In our present attempt to go beyond the mean field, a direct $f$--$f$ hopping amplitude $t^f$ is variationally generated, hence we do not need to include it to get sensible physical results. In particular,  
for moderate values of $U$, the generated nearest neighbor $f$--$f$ hopping has an opposite sign to its $c$--$c$ counterpart $t$, which is convenient to open an indirect gap. In fact, the best situation to open an hybridization gap between two overlapping bands 
occurs when they cross with opposite slopes, which is exactly what our variational wave function does to minimize the energy. The value of the indirect gap $\Delta_g$ that we find is shown as a color plot in the phase diagram of Fig. \ref{fig1}(a).

\begin{center}\vspace{-0.2cm}
  \begin{figure}[t]  \vspace{-0.2cm} 
\includegraphics[width=0.5\textwidth]{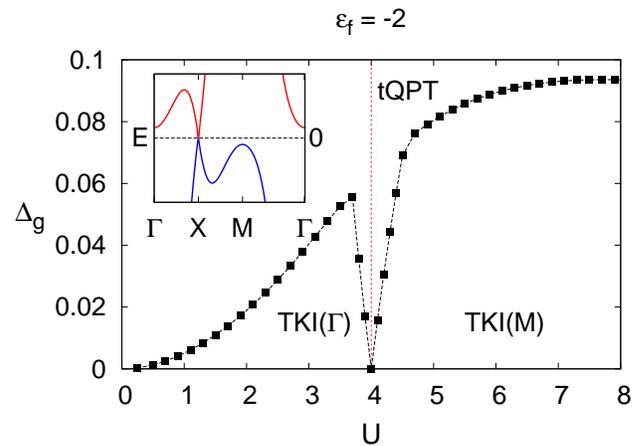}   
   \caption{Evolution of the indirect gap $\Delta_g$ with increasing $U$ across the topological quantum phase transition (tQPT) between TKI($\Gamma$) and TKI(M) at $\varepsilon_f=-2$. The inset shows the low energy band structure at the topological phase transition where both direct and indirect gaps close, which indicates that the TKI($\Gamma$) and TKI(M) are phases not adiabatically connected.  \vspace{-0.2cm}} 
  \label{fig2}
  \end{figure} 
 \end{center}\vspace{-0.7cm}

In terms of the topologically distinct phases in 
the present lattice geometry \cite{Zaanen2013,Assaad2013}, 
we find [see Fig. \ref{fig1}(a)]  topologically trivial insulators (TTIs) 
as well as topologically nontrivial Kondo insulating phases: TKI($\Gamma$) 
with the parities $\delta_i=(-1,1,1,1)$  
and TKI(M) with $\delta_i=(-1,1,-1,-1)$, where $i\in\{\Gamma,{\rm M},{\rm X},{\rm Y}\}$. The exemplary low energy band structures 
for TKI($\Gamma$) and TKI(M) are drawn in Figs.~\ref{fig1}(c) and \ref{fig1}(d). 
The topological phase transition between these two phases 
significantly influences the behavior 
of the indirect gap.

The largest values of the gap are attained 
above $U\sim 4$ and are roughly 
enclosed within the $f$ isovalent regions with $n_f\simeq 0.9$ and 
$n_f\simeq 1.1$  [cf. Fig.\ref{fig1}(a)], where $n_f$ is 
the average $f$ electron number per lattice site. 
The intuitive expectation that the effect
of correlations is more pronounced the closer the $f$ orbital is at 
the half-filling thus 
fails in the vicinity of the TKI($\Gamma$)--TKI(M) 
transition for $n_f\simeq1$, marked with a solid gray 
line in Fig.~\ref{fig1}(a).  
This topological phase transition enforces the closure of 
the band gap (both direct and indirect) \cite{Assaad2013} and 
leads to its rapid decrease 
in its neighborhood (see Fig. \ref{fig2}). 
We note that, even though the behavior of the gap versus $U$ at $\varepsilon=-2$ (Fig.~\ref{fig2}) is different from the DMFT results of Ref.~\cite{Assaad2013}, not surprisingly since we do not 
include any finite $f$--$f$ hopping, nevertheless the value $U\simeq4$
for the topological transition is coincident.   

   \begin{center}\vspace{-0.2cm}
  \begin{figure}[t]
   \includegraphics[width=0.48\textwidth]{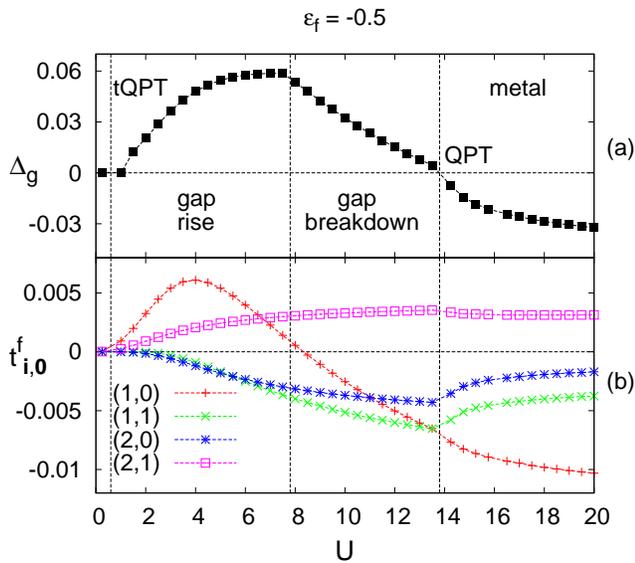}        
    \caption{Evolution vs $U$ of (a) the indirect gap $\Delta_g$; (b) the variationally generated $f$--$f$ hopping amplitudes $t^f$. Three regimes are identified: (i) initial rise of the gap; (ii) its collapse associated with the sign change of the nearest neighbor [$\g r=(1,0)$] $t^f$; (iii)  
the metallic state characterized by the negative value of $\Delta_g$, namely by overlapping valence and conduction bands, after a quantum phase transition (QPT). 
We also mark the topological quantum phase transition (tQPT) for small $U\simeq0.5$.
}
  \label{fig3}
  \end{figure} 
 \end{center}\vspace{-0.7cm}

 In Fig. \ref{fig1}(b), we show the phase diagram at 
larger values of the Coulomb repulsion. Here and for 
$\varepsilon_f \gtrsim-2$, upon increasing $U$, the system undergoes  a transition from a 
TKI to a metal [see Fig.\ref{fig1}(e) for an exemplary low energy 
band structure]. On the contrary, for $\varepsilon_f \lesssim-2$, we find an orbital-selective Mott state where 
the $f$ occupancy is pinned to $n_f=1$ with vanishingly small fluctuations, $d_f\to 0$. 

In Fig.~\ref{fig3}(a) we plot the evolution of $\Delta_g$  with increasing $U$ for $\varepsilon_f=-0.5$.
In this plot we have singled out three stages: ({\it i}) the initial rise of the gap, ({\it ii}) 
its subsequent drop till ({\it iii}) its sign change, which signals that the two bands now overlap, yielding 
a metallic behavior. 
We also mark the topological transition between TKI($\Gamma$) and TKI(M) at small $U\simeq0.5$. It seems that the crucial factor leading to the nonmonotonic gap behavior is just  
the effective $f$--$f$ hopping that is variationally generated. In  
Fig.~\ref{fig3}(b), we show the $U$ dependence of  the $f$--$f$ hopping amplitudes, $t^f_{\g i, \g 0}$, with different $\g i=(n,m)$. We observe  
that the downturn of $\Delta_g$ occurs close to the value of $U$ at which the nearest-neighbor 
hopping, i.e., $\g i = (1,0)$, changes sign, from being opposite to the $c$--$c$ hopping to being concordant. However, the sign change is still not enough to close the gap, since in the meantime sizable further-neighbor hopping has been generated: second, $\g i=(1,1)$, third, $\g i=(2,0)$, and fourth neighbor,    
$\g i=(2,1)$, ones. Eventually, 
when the negative nearest-neighbor hopping overwhelms the others, 
the gap closes.  

We can attempt to rationalize the observed nontrivial behavior of the direct $f$--$f$ hopping by simple arguments.  Given that the Hamiltonian, Eq.~\eqn{ALM}, lacks such hopping process, the expectation value  $\langle f^\dagger_i f^\dagga_j\rangle$ between nearest-neighbor sites on the true ground state is finite only because of $c$--$f$ hybridization and has an opposite sign to $\langle c^\dagger_i c^\dagga_j\rangle$ for any $U$. Since the latter hampers the $f$ electron motion, it must reduce the value of 
$\left|\langle f^\dagger_i f^\dagga_j\rangle\right|$ with respect to the noninteracting $U=0$ case. In a variational approach such as ours, based on a wave function obtained as the ground state of an auxiliary noninteracting Hamiltonian $\mathcal{H}^\text{eff}$ modified by the action of local operators $\mathcal{P}_{\g i}$, this reduction can be attained in two ways: by lowering the $c$--$f$ hybridization $V^\text{eff}$ in $\mathcal{H}^\text{eff}$ and/or by generating a nearest-neighbor $f$--$f$ hopping $t^f$ of the same sign as the $c$--$c$ one, i.e., negative in our case. When $U$ is large, the optimized Hamiltonian $\mathcal{H}^\text{eff}$ indeed comprises a finite $t^f<0$, which, in addition to the lowering of $V^{\rm eff}$, allows one to reduce $\left|\langle f^\dagger_i f^\dagga_j\rangle\right|$. 
This result agrees with more accurate variational approaches in the large-$U$ limit of the Kondo lattice 
model \cite{Fabrizio2013}. 
On the contrary, for small to intermediate values of $U$, the system prefers to generate a positive $t^f>0$, which, as we mentioned, favors the opening of a hybridization gap.

{\it Summary.} 
The results we have obtained might be relevant 
to known Kondo semiconductors \cite{Riseborough2000}
and are promising in view of possible topologically 
nontrivial states. 
The wealth of phases that we find, ranging from nontopological to 
topological insulators and metals, are indeed observed in different 
compounds \cite{Menth1969,Hundley1990,Takabatake1990,Malik1991,Okamura1998,Takabatake2007}.
Moreover, the nonmonotonic behavior of the indirect gap with 
respect to the strength of $U$ brings to mind the variety of responses 
observed under pressure in Kondo semiconductors, 
ranging from a gap decrease in \sm\ \cite{Bauer2003, Fisk2016} and in 
CeRhAs \cite{Yoshii1996}, to its increase in \cebipt\ \cite{Cooley1997} 
and in \cns\ \cite{Hiraoka1994}, and finally to the nonmonotonic evolution 
as seen in \crs\ \cite{Fisk1996}. We also note that the semimetal phase, which appears once the indirect gap closes and still possesses topologically nontrivial properties, mimics exactly the physical scenario proposed 
for \cns\ \cite{Coleman2010,Coleman2016} to explain its intriguing properties \cite{Sera1997,Terashima2002,Slebarski2009,Coleman2016}. 
The precise sequence and topological properties of the phases that we 
find must evidently depend on our choice of model and lattice geometry. 
However, it is encouraging that the same method
gives access to all those phases, 
including topologically nontrivial ones, especially in view of applications to more 
realistic modeling.

In summary we have shown that the Anderson lattice model Eq.~\eqn{ALM} can support a topological insulating state stabilized purely by correlations. In particular, we found variationally that the impurity self-energy acquires the right momentum dependence to open a gap, despite the fact that, from the start, the model does not include any direct hopping between the $f$ orbitals. It is now worth comparing our results with recent ones on \sm\ based on a combined LDA plus Gutzwiller technique \cite{Dai2013}. Within LDA the spin-orbit split  
$f$ orbitals give rise to narrow bands of width $W\!\sim\! 0.5\!$ eV around the Fermi level with a semiconducting behavior characterized by a direct gap $\Delta\!\sim\!15\!$~meV. The inclusion of local Gutzwiller correlations 
brings about a sizable reduction of quasiparticle weight, $z\!\sim\!0.18$, which reduces the $j\!=\!5/2$ 
bandwidth to $W_*\!\sim\! z\,W\sim 0.1\!$~eV, the $j\!=\!7/2$ $f$ bands being pushed up to 4.0~eV above the Fermi level \cite{Dai2013}. 
Despite such a small value of $z$, the semiconducting behavior survives local correlations. The direct gap is almost unchanged, although a smaller indirect gap of $\sim \!10\!$~meV emerges. Although this result is well in accordance with ours, the mechanism that stabilises the gap might at first glance appear to be different, since in the case of Ref.~\cite{Dai2013} the LDA band structure is already semiconducting, which, as we mentioned, implies a finite $f$-$f$ hopping. However, since LDA already includes some correlation effects, it is not clear whether the $f$-$f$ hopping is a one- or many-body effect. The fact that the gap is almost unaffected by the reduction of quasiparticle weight that would renormalize down any one-body hopping term, might suggest that the LDA 
$f$ dispersion already includes nonlocal correlation effects, thus in agreement with our calculations. Further investigations would be desirable to assess this issue.

{\it Acknowledgements.} MMW is grateful for discussions with A. Amaricci, and greatly acknowledges the support 
from Polish Ministry of Science and Higher Education under 
the ``Mobilno\'s\'c Plus'' program, Agreement No. 1265/MOB/IV/2015/0.


\newpage \clearpage

\renewcommand{\thepage}{S\arabic{page}} 
\renewcommand{\thesection}{S.\Roman{section}}  
\renewcommand{\thetable}{S\arabic{table}}  
\renewcommand{\thefigure}{S\arabic{figure}} 
\renewcommand{\theequation}{S\arabic{equation}} 
\setcounter{page}{1}
\setcounter{equation}{0}
\setcounter{figure}{0}
\setcounter{section}{0}

\begin{widetext}
\begin{center}
{\large \bf Supplemental Material to the article: \\ ``Many-body breakdown of indirect gap in topological Kondo insulators''} 
\end{center}
\end{widetext}

\section{Details of the diagrammatic expansion for the Gutzwiller wave function technique}
\subsection{Formal expansion}
The  variational analysis of the topological 
Anderson lattice Hamiltonian, including the non-local 
effects of the onsite interaction, begins with the formulation of the exact 
form of the expectation value with the full Gutzwiller wave function (GWF) 
of any operator $\mathcal{O}_I$ 
that involves sites $I=\big\{\g i_1,\g i_2, \dots,\g i_M\big \}$,
\begin{equation}
\begin{split}
 \big\langle \, &\mathcal{O}_{I}\,\big\rangle_G = \frac{1}{\langle\psi_G|\psi_G\rangle}
 \Big\langle \,{\mathcal{O}}^G_{I}\;\prod_{\g l\not \in I}\,  
 \mathcal{P}_{\g l}^2 \, \Big \rangle_0    \\
 &= \frac{1}{\langle\psi_G|\psi_G\rangle}\sum_{k=0}^\infty\, \fract{x^k}{k!}\, \sum_{\g l_1,...,\g l_k}\!\!\!\!{}^{'}
 \Big\langle \,{\mathcal{O}}^G_{I}\, d_{\g l_1}\,d_{\g l_2}\dots d_{\g l_k}\,\Big
\rangle_0\; ,
 \label{expeS}
\end{split}
 \end{equation}
where  $\mathcal{O}^G_{I}\equiv \prod_{i\in I}\,\mathcal{P}_{\g i}^\dagger\;\mathcal{O}_I\; 
\prod_{j\in I}\,\mathcal{P}_{\g j}^\dagga$. 
The expectation values in Eq.~(\ref{expeS}) are evaluated by means of the 
Wick's theorem. They can be visualized by a sum  of diagrams connecting sites 
in the real space with lines representing single particle density matrices,
\begin{equation}
\begin{split}
 C_{\g{l},\g{l'}}^{ff} &\equiv \langle  f_{\g{l}\sigma}^\dagger
  f_{\g{l'}\sigma}^\dagga \rangle_0,\\
C_{\g{l},\g{l'} }^{fc}&\equiv \langle  c_{\g{l}\sigma}^\dagger
  f_{\g{l'}-\sigma}^\dagga \rangle_0,\\
C_{\g{l},\g{l'}}^{cc}&\equiv \langle  c_{\g{l}\sigma}^\dagger
  c_{\g{l'}\sigma}^\dagga \rangle_0,
 \label{linesS}
 \end{split}
 \end{equation}
where    $\g l,\g l'$ are 
lattice indices and $\sigma$ denotes (pseudo)spin index for 
$c$ ($f$) electrons. In order to evaluate the expectation value of the topological 
Anderson lattice Hamiltonian 
first we derive the expressions for the following projected operators,  
\begin{equation}
\begin{split}
\mathcal{ P}_{\g i}  d_{\g i} \mathcal{ P}_{\g i}&=\lambda_d^2[2n_{0f}\ n^{\rm  HF}_{\g i}+(1-xd_0)\ d_{\g i} +n_{0f}^2\ \mathcal{ P}_{\g i}^2], \vspace{2pt} \\
\mathcal{ P}_{\g i}  n^f_{\g i\sigma} \mathcal{ P}_{\g i}&=(1+xm)\ n^{\rm  HF}_{\g i}+\gamma\ d_{\g i}+n_{0f}\ \mathcal{ P}_{\g i}^2,  \vspace{2pt} \\
\mathcal{ P}_{\g i}  f_{\g i\sigma}^{(\dagger)} \mathcal{ P}_{\g i}&=\alpha\ f_{\g i\sigma}^{(\dagger)}+
\beta\ f_{\g i\sigma}^{(\dagger)} n^{\rm  HF}_{\g i}, \label{A11S}
\end{split}
\end{equation}
where additionally we define
\begin{equation}
\begin{split}
 n^{\rm  HF}_{\g i\sigma}&\equiv n^f_{\g i\sigma} - \langle n^f_{\g i\sigma}\rangle_0 = n^{\rm  HF}_{\g i-\sigma} \equiv  n^{\rm  HF}_{\g i}\\
n_{0f}&\equiv\langle n^f_{\g i\sigma}\rangle_0 \\
\beta&\equiv\lambda_{s}( \lambda_d-\lambda_0),\\
\alpha&\equiv\lambda_s \lambda_0+\beta n_{0f},\\
\gamma&\equiv x(1-2n_{0f}),\\
 m&\equiv n_{0f}(1-n_{0f}),\label{A4S}
\end{split}
\end{equation}
where superscript HF stands for the Hartree-Fock form of the operator. 
Here, the parameters $\{\lambda_0,\lambda_s,\lambda_d\}$ have the following meaning.
The general, diagonal Gutzwiller correlator acting on the $f$ 
electrons can be written as,
\begin{equation}
 \mathcal{ P}_{\g i}=\sum_\Gamma \lambda_\Gamma\mid\Gamma\rangle_{\g i}\langle\Gamma\mid_{\g i},
\end{equation}
with variational parameters 
$\lambda_\Gamma \in \{\lambda_0,\lambda_\uparrow,\lambda_\downarrow,\lambda_d\}$ that 
characterize the occupation probabilities for the four possible atomic Fock $f$-states 
${\mid\Gamma\rangle_{\g i}\in\{\mid 0\rangle_{\g i},\mid\uparrow\rangle_{\g i},
\mid\downarrow\rangle_{\g i},\mid\uparrow\downarrow\rangle_{\g i}\}}$.
By constraining the correlator with Eq.~(2) (main manuscript) we obtain
\begin{equation}
 \begin{split}
 \lambda_0^2&=1+xn_{0f}^2,\\
\lambda_\sigma^2=\lambda_{- \sigma}^2\equiv\lambda_s^2&=1-xn_{0f}(1-n_{0f}),\\
\lambda_d^2&=1+x(1-n_{0f})^2. \label{ddS}
\end{split}
\end{equation}

The diagrammatic sums for the operators resulting from (\ref{A11S}) are expressed as 
\begin{equation}
S=\sum_{k=0}^\infty \frac{x^k}{k!}S(k),                                                                                                                                                                                                                                                                                                                                                               \end{equation}
where  $ S\in\{ T^{(1,1)}_{\g i,\g i+\g x}, 
T^{(1,3)}_{\g i,\g i+\g x},I^{(4)},I^{(2)}\}$,
and the k-th order contributions $S(k)$ read,
\begin{equation}
\begin{split}
T^{(1,1[3])}_{\g i,\g i+\g x}(k)&\equiv\sum_{\g l_1,...,\g l_k}\langle  c_{\g i\sigma}^\dagger
[ n^{\rm HF}_{\g i,\sigma}]  f_{\g i+\g x,-\sigma} d_{\g l_1,...,\g l_k}\rangle_0^c\\
I^{(4)}(k)&\equiv\sum_{\g l_1,...,\g l_k}\langle  d_{\g i} 
 d_{\g l_1,...,\g l_k}\rangle_0^c,\\
I^{(2)}(k)&\equiv\sum_{\g l_1,...,\g l_k}\langle  n_{\g i}^{\rm HF} d_{\g l_1,...,\g l_k}\rangle_0^c.
\end{split}
\end{equation}
Here superscript $c$ denotes only fully connected diagrams \cite{buenemann2012S}, 
because, with the help of the linked-cluster theorem waiving the summation restriction, 
GWF norm cancels out all the disconnected diagrams.

In the Eqs. (\ref{A11S}) we have expressed the projected operators by their HF or relative forms as it significantly speeds up the 
convergence of the numerical results concerning the summation of the  diagrams \cite{buenemann2012S}.
 It is attributed to the fact that by such a construction, all the two-operator averages for a single site and $f$-orbital 
(so-called {\it Hartree bubbles}) automatically vanish.
 
\subsection{Conservation of the number of electrons}
Before the explicit evaluation of the expectation value of the topological Anderson lattice Hamiltonian with the GWF one issue,
particularly important for the modelling of Kondo insulators,  
still needs to be resolved. 
Due to the fact that the Gutzwiller correlator has a 
diagonal form and acts only on the $f$ degrees of freedom it commutes 
not only with the operator of the total number of particles but also with those of $f$ and $c$ number of electrons separately. 
This yields following equities,
\begin{equation}
\begin{split}
  \langle  c_{\g{i}\sigma}^\dagger  c^\dagga_{\g{i}\sigma} \rangle_G&=\langle  c_{\g{i}\sigma}^\dagger  c^\dagga_{\g{i}\sigma} \rangle_0,\\
\langle  f_{\g{i}\sigma}^\dagger  f^\dagga_{\g{i}\sigma} \rangle_G&=
\langle  f_{\g{i}\sigma}^\dagger  f^\dagga_{\g{i}\sigma} \rangle_0.
\end{split}
\end{equation}
It is not automatically satisfied in our approach 
because of the real-space and the order cut-offs introduced \cite{Kaczmarczyk2015S}. However, 
it can be straightforwardly enforced by the construction.
Due to the odd hybridization, introducing symmetric 
cancellation of the diagrams in the real space, the 
averages of the pairs of $c$ operators for all $\g i$~and $\g j$, are 
the same when taken with either $|\psi_0\rangle$ or $|\psi_G\rangle$,  
\begin{equation}
\langle c_{\g{i}\sigma}^\dagger  c^\dagga_{\g{j}\sigma} \rangle_G=
\langle  c_{\g{i}\sigma}^\dagger  c^\dagga_{\g{j}\sigma} \rangle_0.
\end{equation}
Therefore, what remains, is to ensure that ${\langle  f_{\g{i}\sigma}^\dagger  f_{\g{i}\sigma}^\dagga \rangle_G=
\langle  f_{\g{i}\sigma}^\dagger  f^\dagga_{\g{i}\sigma} \rangle_0}$. The {\it correlated} 
(averaged with the GWF) number of $f$ electrons can be rewritten in the following form (cf. Eq.~(\ref{A11S})),
\begin{equation}
\begin{gathered}
 \langle  f_{\g{i}\sigma}^\dagger  f^\dagga_{\g{i}\sigma} \rangle_G= n_{0f}+(1+xm)I^{(2
)}+\gamma I^{(4)}.
\label{n0nfS}
\end{gathered}
\end{equation}
The conservation of the total number of particles can be now enforced by the construction by 
imposing the cancellation 
of the sum of the two last terms  
in Eq. (\ref{n0nfS}) \cite{buenemann2012S}, which reduces to the following relation
\begin{equation}
I^{(2)}=\frac{-\gamma}{1+xm}I^{(4)}. 
\end{equation}
As a result we calculate diagrammatically only $I^{(4)}$, which due to its structure introduces less 
error connected with the real-space cut-off than $I^{(2)}$.\cite{Kaczmarczyk2015S}

\subsection{Expectation value of the Hamiltonian and the derivation of its single-particle effective correspondant}
The resulting expectation value of the topological Anderson lattice Hamiltonian 
with the GWF can be readily expressed as following ($\sigma$ and lattice summations are already executed),
\widetext
\begin{equation}
\begin{split}
\frac{\langle{\mathcal{H}}\rangle_G}{L} = 8 
 t C^{cc}_{\g{i},\g{i}+\g{x}}-2\mu C^{cc}_{\g{i},\g{i}}-2(\mu-\varepsilon_f)n_{0f} 
 +U\lambda_d^2\Big(n_{0f}^2+\frac{\lambda_s^2\lambda_0^2}{1+mx} I^{(4)}\Big) 
 +16 V \Big(\alpha T^{(1,1)}_{\g i,\g i+\g x} +\beta T^{(1,3)}_{\g i,\g i+\g x}\Big),
\end{split}
\end{equation}
where L denotes the number of the lattice sites and $\g x$ is a vector in the $\g x$-direction with 
the length of the lattice constant. Calculated diagrammatically
$\langle{\mathcal{H}}\rangle_G$ is first optimized with respect to the
variational parameter $x$. Then the optimization of Slater determinant is proceeded what leads  to the construction of 
the effective single-particle Hamiltonian $\mathcal{ H}^{\rm eff}$. Effectively, 
it reduces to fulfilling the condition that its optimal
expectation value with $|\psi_0\rangle$  
is coincident with that calculated diagrammatically for $\mathcal{H}$ with
 $|\psi_G\rangle$,  
 \begin{equation}
\begin{gathered}
\delta\langle \mathcal{ H}^{\rm eff}\rangle_0
(C_{\g{l},\g{l'}}^{\alpha,\beta},n_{0f})= 
\delta\langle \mathcal{ H}\rangle_{\rm G} (C_{\g{l},\g{l'}}^{\alpha,\beta},n_{0f}) 
=\sum_{\g l,\g{l'},\alpha,\beta} \Big{(}\frac{\partial \langle \mathcal{ H}\rangle_{\rm G}}
{\partial C_{\g{l},\g{l'}}^{\alpha,\beta}}\delta
C_{\g{l},\g{l'}}^{\alpha,\beta} 
 +\frac{\partial \langle \mathcal{ H}\rangle_{\rm G}}{\partial n_{0f}}\delta
n_{0f}\Big),
\end{gathered}
\end{equation} 
where $\alpha,\beta \in\{f,c\}$. This yields following structure of  $\mathcal{ H}^{\rm eff}$ presented in the main text of the manuscript,
\begin{eqnarray}
\mathcal{H}^{\rm eff}\!\!\!\!&=& \sum_{\g i,\g j}
 t_{\g i\g j} \, \hat c_{\g i }^\dagger\, \hat c_{\g j }^\dagga
+\sum_{\g i,\g j}t_{\g i \g j}^{f}\, \hat f_{\g i }^\dagger \hat f_{\g j}^\dagga\nonumber \\ 
&&\!\!\!\!\!\!\!\!\!+\!\!\!\sum_{\langle\g i,\g j\rangle,\alpha=x,y}\bigg[\,iV_{\g i \g j}^{\rm eff}\,\Big(\hat c_{\g i}^\dagger\,\sigma_\alpha\,
\hat f_{\g j}^\dagga +\hat f_{\g i}^\dagger\,\sigma_\alpha\, 
\hat c_{\g j}^\dagga\Big) +{\rm H.c.}\,\bigg] + \sum_{\langle\langle\g i,\g j\rangle\rangle}\bigg( iV_{\g i \g j}^{\rm eff}\Big[\hat c_{\g i}^\dagger\big(\sigma_x-a\sigma_y\big)
\hat f_{\g j}+\hat f_{\g i}^\dagger\big(\sigma_x-a\sigma_y\big)
\hat c_{\g j}\Big]+\rm{H.c.}\bigg)\nonumber\\
&\equiv& \sum_{\mathbf{k}}\, \hat{\Psi}_\mathbf{k}^\dagger\, \mathcal{\hat{H}}^\text{eff}(\mathbf{k})\,
\hat{\Psi}_\mathbf{k}^\dagga,\label{neeS}
\end{eqnarray}
The  effective microscopic parameters
determining $\mathcal{ H}^{\rm eff}$ read (note that the $c$-electron hopping remains unchanged)   
\begin{equation}
\begin{split}
t_{\g i\g j}^{f }\Big|_{\g i\neq \g j}&= \frac{\partial \langle\mathcal{H}\rangle_G}
 {\partial F_{\g{i},\g{j}}}, \ \ \ \ \ \ \ \ \
 V_{\langle \g i\g j\rangle}^{\rm eff}= \frac{\partial \langle\mathcal{H}\rangle_G}
 {i\ \partial  W_{\langle \g i,\g j\rangle_x}},\\
t_{\g i\g i}^{f}&= 
 \frac{\partial \langle\mathcal{H}\rangle_G}
 {\partial n_{0f}},   \ \ \ \ \ \
 V_{\langle\langle \g i,\g j\rangle\rangle}^{\rm eff} = \frac{\partial \langle\mathcal{H}\rangle_G}
 {(i-1) \partial  W_{\langle\langle \g i,\g j\rangle\rangle_{x=y}}}.\label{teffS}
\end{split}
\end{equation}
 \endwidetext
\subsection{General scheme}
Finding equilibrium groundstate can be described in the following steps:\\
1. Diagrammatic  evaluation of $\langle\mathcal{ H}\rangle_{\rm G}$ for the chosen
$|\psi_0\rangle$.\\
2. Minimization of $\langle\mathcal{ H}\rangle_{\rm G}$ with respect to $x$. \\
3. Determination of the single-particle Hamiltonian $\mathcal{ H}^{\rm eff}$ by 
optimization of $\langle\mathcal{ H}\rangle_{\rm G}$ with $|\psi_0\rangle$.\\ 
4. Determination the ground state, $|\psi_0'\rangle$ of  $\mathcal{
H}^{\rm eff}$. \\
5. Repeating  steps 1-4 in a self-consistent loop until a satisfactory
convergence for $|\psi_0\rangle=|\psi_0'\rangle$ is reached.
 
 \section{Convergence with respect to the order of expansion}
\label{sec2S}
In Fig. \ref{fig2S} we present the resulting indirect gap value for 
$\varepsilon=-0.5$ (cf. Fig. 3(a) - main manuscript) for different orders of the diagrammatic expansion.
We do not show results for $k=0$ as on the recovered there mean-field 
level the topological Anderson lattice model with initially non-dispersive $f$-states 
is capable to describe only semi-metallic state with a zero gap. 
In the subsequent orders, $k>0$ the nontrivial behavior of a gap can
be seen. Starting from the second order, $k\geq2$ the metallic state 
appears in the strongly correlated regime $U\gtrsim10$. 
The practically overlaping values of the gap 
for  $k=4$ and $k=5$ indicate that at this level of truncation of the
 order the satisfactory convergence is reached. Hence all results in the main manuscript
 are presented for $k=4$. \ \ \ \ \  \ \phantom{a}

 \begin{center}\vspace{-1cm}
  \begin{figure}[h]   
\includegraphics[width=0.47  \textwidth]{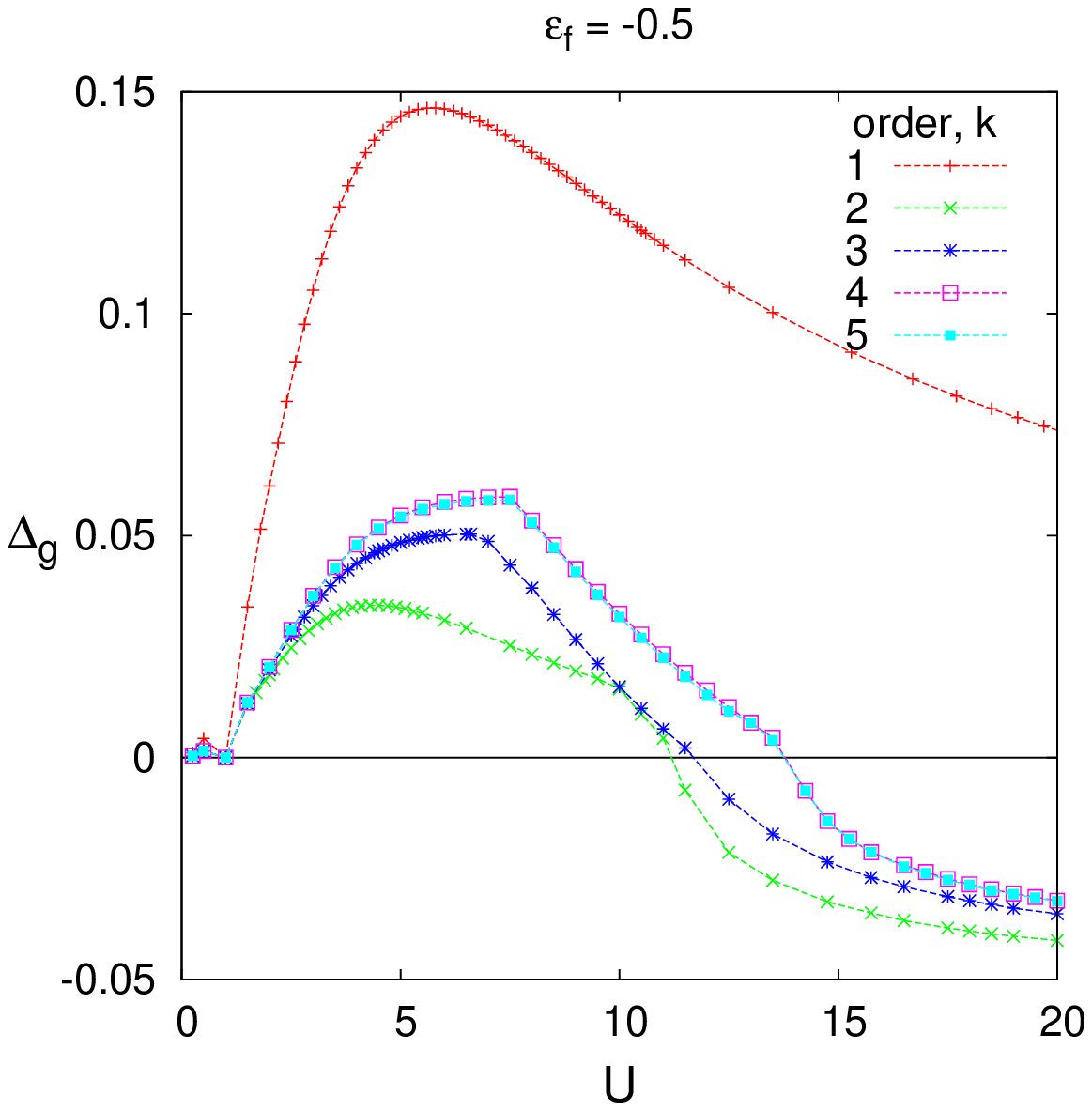}   
   \caption{Value of the indirect gap, $\Delta_g$ with respect to the interaction strength
$U$. The negative value of the gap denotes the range of the mutual 
energy overlap between the upper and the lower hybridization bands in the metallic state. 
The points for $k=4$ and $k=5$ practically overlap indicating that satisfactory convergence with respect to order is reached.} 
  \label{fig2S}
  \end{figure} 
 \end{center}

\end{document}